\documentclass[aps,prl,twocolumn,nofootinbib,groupedaddress,amsfonts,floatfix]{revtex4}

\usepackage{graphicx,amsmath,amssymb,amstext}
\usepackage{amssymb,amsbsy,amsfonts,amsthm,hyperref}
\usepackage{lscape,placeins}
\usepackage{url}
\makeatletter
\g@addto@macro{\UrlBreaks}{\UrlOrds}
\makeatother

\newcommand{\gsim}{\;\mbox{\raisebox{-0.5ex}{$\stackrel{>}{\scriptstyle{\sim}}$}}\;}
\newcommand{\lsim}{\;\mbox{\raisebox{-0.5ex}{$\stackrel{<}{\scriptstyle{\sim}}$}}\;}

\usepackage{color}
\usepackage{ifthen}
\newboolean{editorial}
\setboolean{editorial}{true}
\newcommand{\editorial}[2]{\ifthenelse{\boolean{editorial}}{\textcolor{red}{[\textsf{\textbf{{#1}}}: }\textcolor{blue}{\textsf{{#2}}}\textcolor{red}{]}}{}}

\begin{document}

\title{Could 1I/'Oumuamua be macroscopic dark matter?}

\author{David Cyncynates${}^{3}$}
\author{Emanuela Dimastrogiovanni${}^{1,2}$}
\author{Saurabh Kumar${}^{1}$}
\author{Jagjit Sidhu${}^{1}$}
\author{Glenn D. Starkman${}^{1}$}

\affiliation{${}^1$CERCA/ISO/Department of Physics, Case Western Reserve University, 10900 Euclid Avenue, Cleveland, OH 44106}
\affiliation{${}^2$Perimeter Institute for Theoretical Physics, 31 Caroline St. N., Waterloo, ON, N2L 2Y5, Canada}
\affiliation{${}^3$Stanford University, Stanford, California 94305, USA}

\begin{abstract}
1I/'Oumuamua, formerly known as A/2017 U1, 
is a sizable body currently passing through the solar system.
It is generally considered to be a rocky asteroid-like object
that came from another planetary system in the Milky Way.  
We point out that 1I/'Oumuamua may instead be a chunk of 
dark matter, a ``macro,''
possibly as massive as $10^{25}$g if it is of nuclear density. 
If so, then its passage will have caused measurable deviations in the 
orbits of Mercury, the Earth and Moon.
\end{abstract}

\maketitle

On October 19, 2017, an ``unusual object'' 
was observed to be passing through the solar system. 
Temporarily designated A/2017 U1 and now renamed 1I/'Oumuamua
by the Minor Planet Center (MPC) in Cambridge, Massachusetts, 
it was discovered by R. Weryk using the Pan-STARRS 1 telescope,
and subsequently identified by him in the archived images of Oct 18.
1I/'Oumuamua appeared to come from within $6^\circ$ of the solar apex --
the direction that the Sun is moving (at about $20$ km/s) 
through the solar neighborhood --  
with a hyperbolic excess velocity 
of approximately $26$ km/s with respect to the Sun \cite{SkyandTelescopeA-2017U1}.
This direction is almost directly ``above'' the ecliptic, 
so 1I/'Oumuamua did not have any close encounters with a solar-system planet before 
reaching perihelion inside Mercury's orbit on Sept. 9, 
slingshotting under the Sun, and passing within about 24 million km of the Earth
on its way back out of the Solar system.

K. Meech \cite{MeechK} 
``reported that in a very deep stacked image, 
obtained with the VLT, this object appears completely stellar.''
In other words, 1I/'Oumuamua was too small to be resolved. 
It appears to have a featureless red spectrum.
B. Gray calculated \cite{SkyandTelescopeA-2017U1}
that 1I/'Oumuamua would have a diameter of about 160 meters 
if it were a rock with a surface reflectivity of 10\%. 
Explanations for 1I/'Oumuamua have focused on the possibility that 
it is an interstellar asteroid ejected 
from an extrasolar planetary system \cite{NASAannounemcnet}.

An alternative  is that 1I/'Oumuamua is not any ordinary rock, 
but a macroscopic chunk of dark matter,  a ``macro'' \cite{Jacobs:2014yca}.
Contrary to widely held misconceptions, dark matter need not be in the form of 
weakly interacting elementary particles, 
but might instead be found in much larger pieces 
with masses best measured in grams or kilograms,
and cross-sections best measured in cm$^2$.
Specific candidates include 
primordial black holes \cite{ZeldovichNovikov1967,Carr:1974nx}, 
strange quark or baryonic matter \cite{Witten:1984rs,Lynn:1989xb,Lynn:2010aa}, 
and other speculative approximately nuclear-density 
Standard-Model or Beyond-the-Standard-Model objects 
(nuclearites \cite{DeRujula:1984axn}, 
quark nuggets \cite{Gorham:2012aa,Zhitnitsky:2016aa}, 
CUDOS \cite{Rafelski:2013ab}, {\it etc.}).

Given that 1I/'Oumuamua was seen to reflect sunlight, 
we can eliminate the possibility that it is a primordial black hole.
Strange baryonic matter, however, will reflect light, 
though no {\it ab initio} calculation of its reflectivity or spectrum would be reliable.
Extending the calculations of Gray \cite{SkyandTelescopeA-2017U1},
we can conclude that if 1I/'Oumuamua is spherical it has a  cross-sectional area 
\begin{equation}
\label{sigmaxmin}
\sigma_X \geq \sigma_X^{\mathrm{min}}=2\times10^7{\mathrm{cm}}^2 \,.
\end{equation}
\begin{figure}
	\centering
	\includegraphics[width=\hsize]{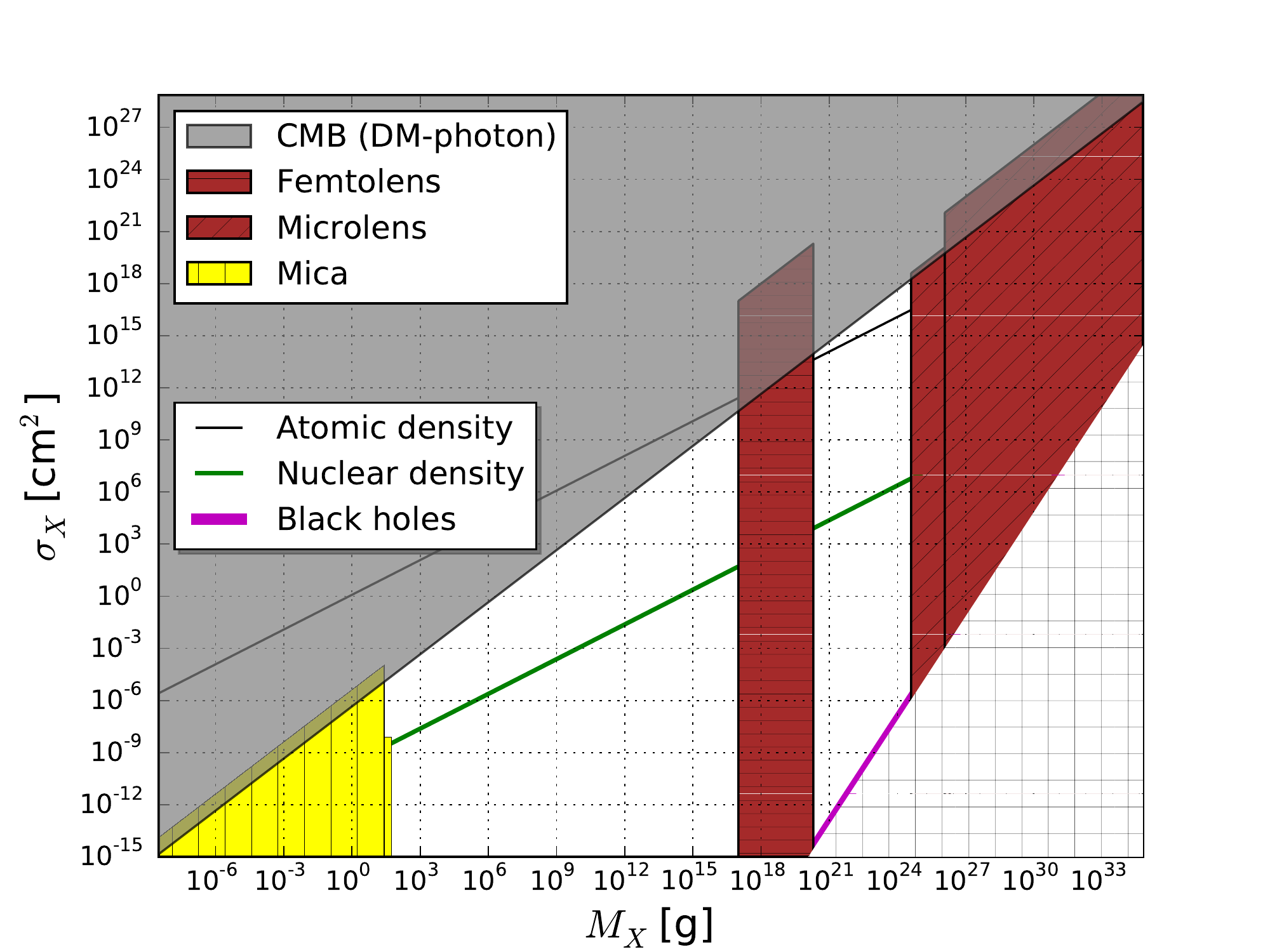}
	\caption{Figure 3 of \cite{Jacobs:2014yca}. 
	Constraints on the Macro cross-section and mass 
	(assuming all Macros have the same mass and size).
	Gravitational-lensing constraints are in red; 
	CMB-inferred constraints are in grey;
	constraints from recovered and etched mica are in yellow.
	The black and green lines correspond to objects of constant density 
	$1  {\mathrm{g/cm^3}}$ and $3.6\times10^{14} {\mathrm{g/cm^3}}$, respectively. 
	Black hole candidates lie on the magenta line; 
	objects below and to the right of that line should not exist.}
	\label{JacobsetalFig3} 
\end{figure}
From figure \ref{JacobsetalFig3} we see that macros with $\sigma_X$
respecting (\ref{sigmaxmin}) are allowed, even at 100\% of the dark matter density
for 
\begin{equation}
3\times10^{13}{\mathrm{g}}\lsim M_X \lsim 10^{17}{\mathrm{g}}
\quad \& \quad
2\times 10^{20}{\mathrm{g}} \lsim 2\times 10^{24}\mathrm{g}.
\end{equation}
This does not quite extend to objects of nuclear density, 
which would have 
$M_X\gsim2\times 10^{25}\hskip 2 pt {\mathrm{g}}=10^{-8}M_{\odot}$.  
However, a distribution of masses that included this is allowed \cite{Kuhnel:2017aa}.

We can estimate, 
how often one would expect a macro in this mass range to enter the inner solar system.
Given a local density of dark matter of $\rho_{DM}\simeq 7\times10^{-25}{\mathrm{g/cm^3}}$,
and a relative velocity for dark matter of $\sim250{\hskip 2 pt \mathrm{km/s}}$,
and taking the ``target'' to be the inner astronomical unit, we find the rate to be
\begin{equation}
\Gamma \simeq 4\times10^{-8} \frac{10^{25}\hskip 2 pt \mathrm{g}}{M_X} {\mathrm{yr}}^{-1} \,.
\end{equation}
We see that if 1I/'Oumuamua is a macro of nuclear density (the high end of the mass range),
then its appearance, within about a decade of us being able to see it with a telescope
like Pan-STARRS, would be quite fortuitous.  
The precise interpretation of this rate however is complicated by the fact that 
although 1I/'Oumuamua was on a hyperbolic orbit, 
it was  moving approximately ten times more slowly
than would be expected for generic halo-dark-matter. 
For an isothermal halo, only a small fraction of the halo dark matter would 
be expected to move this slowly.   

On the other hand,  an object of this
size but mass  $10^{13}$g (the minimum allowed per figure \ref{JacobsetalFig3})
would be expected to enter the inner solar system several times per hour.
This suggests that we could rule out such objects as the dark matter 
by their non-observation.

If 1I/'Oumuamua is indeed dense macroscopic dark matter, 
then we would expect that its passage close to Mercury, the Earth and Moon
would have had measurable gravitational effects on their orbits.
Simple estimates suggest that a passing macro would cause a displacement of
approximately 
\begin{equation}
\Delta x \sim \frac{G_NM_X}{v^2} \simeq 300 \hskip 2pt \mathrm{m} \left(\frac{M_X}{10^{25}\hskip 2 pt \mathrm{g}}\right) 
									\left(\frac{50 \hskip 2 pt \mathrm{km/s}}{v}\right)^2
\end{equation}
in a body's orbit, while altering its velocity by  
\begin{eqnarray}
\Delta v \sim \frac{G_NM_X}{r v} &\simeq& 
	1.4 \times 10^{-3} \hskip 2 pt \mathrm{m/s} \left(\frac{M_X}{10^{25}\hskip 2 pt \mathrm{g}}\right) \\
							&&\times \left(\frac{10^7 \hskip 2 pt \mathrm{km}}{r}\right)\left(\frac{50 \hskip 2 pt \mathrm{km/s}}{v}\right)\,.\nonumber
\end{eqnarray} 
Such displacements should be detectable. 
Preliminary estimates of the effects on orbital semi-major axes 
suggest that these are several times larger.
Meanwhile, 
the tidal force on the extremely well measured Earth-Moon system should displace them 
relative to one another by an amount smaller than $\Delta x$ 
by a factor of approximately the lunar orbital radius divided by the distance of closest approach of the macro. 
This works out to a very detectable 
$\Delta x_\text{Tidal} \simeq 10 \hskip 2 pt \mathrm{m}$.
In an upcoming work we will compare the observed motion of Mercury, 
the Earth, the  Moon and 1I/'Oumuamua 
to their predicted paths, 
in an attempt to measure or place an upper limit on the mass of 1I/'Oumuamua.

The search for high density macros, such as primordial black holes 
or lumps of strange baryonic matter, must continue, 
even if one may have just crossed our paths.
Perhaps in the future we will see a macro soon enough to rendezvous
and measure its size and mass,
detect the seismic signal of a macro's lunar or terrestrial 
	impact \cite{DeRujula:1984axn,Herrin:1995es,Cyncynates:2017aa}
or pick up the  gravitational wave signal \cite{Kuhnel:2017ab}
of macros spiraling into the Milky Way's central black hole.

{\sl Acknowledgments}
We thank David Jacobs, Bryan Lynn and Kellen McGee for  influential and ongoing 
conversations on the detectability and fundamental physics of macros.
SK, JS  and GDS are  supported by a Department of Energy grant DE-SC0009946 to GDS.
ED was supported in part by Perimeter Institute for Theoretical Physics. Research at Perimeter Institute is supported by the Government of Canada through Industry Canada and by the Province of Ontario through the Ministry of Economic Development and Innovation.
\bibliography{references}
 
\end{document}